\documentclass[amsmath, amsfonts, superscriptaddress, showpacs, twocolumn, prl]{revtex4}
\usepackage{graphicx}
\usepackage{epsfig}
\usepackage{bm}
\usepackage{dcolumn}
\usepackage{amsmath}
\usepackage{amssymb}
\usepackage{color}

\newcommand{\dv}{\mathop{\rm div}\nolimits}
\newcommand{\curl}{\mathop{\rm curl}\nolimits}

\begin{document}

\title{Viscous magnetoresistance of correlated electron liquids}

\author{Alex Levchenko}
\affiliation{Department of Physics, University of Wisconsin-Madison, Madison, Wisconsin 53706, USA}

\author{Hong-Yi Xie}
\affiliation{Department of Physics, University of Wisconsin-Madison, Madison, Wisconsin 53706, USA}

\author{A. V. Andreev}
\affiliation{Department of Physics, University of Washington,
Seattle, Washington 98195, USA}

\begin{abstract}
We develop a theory of magnetoresistance of two-dimensional electron systems in a smooth disorder potential in the hydrodynamic regime. Our theory applies to two-dimensional semiconductor structures with strongly correlated carriers when the mean free path due to electron-electron collisions is sufficiently short. The dominant contribution to magnetoresistance arises from the modification of the flow pattern by the Lorentz force, rather than the magnetic field dependence of the kinetic coefficients of the electron liquid. The resulting magnetoresistance is positive and quadratic at weak fields. Although the resistivity is governed by both viscosity and thermal conductivity of the electron fluid, the magnetoresistance is controlled by the viscosity only. This enables extraction of viscosity of the electron liquid from magnetotransport measurements.
\end{abstract}

\date{February 17, 2017}

\pacs{72.10.-d, 73.43.Qt, 73.63.Hs}

\maketitle

Low temperatue electron transport in metals and semiconductors may often be understood semiclassically. In the classical theory  electrical resistivity is related to the relaxation rate of quasimomentum of electrons, which is determined  by the interplay of electron-impurity, electron-phonon and electron-electron collisions \cite{LL-Kinetics,GL,Ziman}. In the conventional framework based on the Boltzmann equation these scattering processes are described by independent collision integrals, and the resistivity turns out to be proportional to the sum of partial momentum relaxation rates due to each collision type -- the result known as the Matthiesen's rule. In this approximation the momentum conserving electron-electron collisions do not affect the electrical resistivity. However if correlations between different scattering processes are taken into account then momentum conserving electron-electron collisions also affect the resistivity. As noted long ago by Gurzhi~\cite{GurzhiUFN}, the effect of momentum conserving scattering becomes especially important in the hydrodynamic regime, where  the rate of momentum-conserving scattering significantly exceeds the rate of momentum-relaxing processes.

In situations where momentum relaxation occurs at the sample boundary (as in the Poiseuille flow~\cite{Gurzhi_Poiseuille}) or at point-like scattering centers~\cite{Spivak_Hruska} the electron liquid is \emph{isentropic} and the  flow in linear response is Stokesian, with the resistivity being proportional to the shear viscosity of the electron liquid. Experimental support for the Stokes flow in electron transport has been reported in thin Potassium wires~\cite{Potassium_Gurzhi}, and more recently in other metallic systems~\cite{Molenkamp-PRB94,Molenkamp-PRB95,Mackenzie}.

Over the past few years, the role of momentum-conserving scattering and hydrodynamic effects in electrical resistivity of low-dimensional systems was actively studied in the context of modern high mobility nanostructures. This includes equilibration effects in one-dimensional wires \cite{Micklitz,AL,Matveev,Reider,DeGottardi}, high mobility semiconductor heterostructures with strongly correlated carriers, such as $p$- and $n$-doped GaAs and SiGe quantum wells, as well as Si-MOSFETs~\cite{Kravchenko,Spivak}, and graphene  devices \cite{Fuhrer-1,Fuhrer-2,Geim,Kim-1,Kim-2,Muller-Sachdev,Foster-Aleiner,Narozhny,Polini-Viscosity,Briskot,Principi,Levitov-1,Xie,Polini-Nonlocal,Polini-Whirpools,Levitov-2,Levitov-3}. In these  systems electrons move in the presence of a smooth disorder potential with long range correlations. As a result,  the electron liquid in equilibrium is not isentropic, and the hydrodynamic flow becomes non-Stokesian. In the presence of the flow the liquid develops temperature gradients that are \textit{linear} in the current. The resulting resistivity depends not only on the viscosity of the electron liquid but also on its thermal conductivity. In the context of semiconductor systems with strongly correlated carriers, a theory of resistivity in this regime was developed in Ref.~\onlinecite{Andreev}.

Magnetoresistance (MR) in the hydrodynamic regime arises from spin-polarization of the electron liquid and from the orbital effect of the magnetic field. Theory of orbital  magnetoresistance for the Stokes flow   was recently developed in Ref.~\onlinecite{Alekseev}. In this case MR arises from the magnetic field dependence of the viscosity of the liquid and is controlled by the ratio of the electron mean free path (limited by electron-electron scattering) to the cyclotron radius  $R_c$.

For the hydrodynamic flow in a long-ranged disorder potential the theory of orbital MR   has not been developed. We develop such a theory in the present paper.  We show that although the zero field resistivity depends on both viscosity and thermal conductivity of the electron liquid, MR depends only on the shear viscosity. Furthermore,  the orbital MR becomes appreciable at much weaker fields $H$, at which the viscosity and thermal conductivity may be assumed independent of $H$. In contrast to the approach of Ref.~\onlinecite{Andreev}, which is based on entropy production, we obtain our results by directly evaluating the drag force exerted by the disorder potential onto the flowing electron liquid. This ``mechanical'' approach is better suited for consideration of magnetoresistance. It also enables us to elucidate the physics of the drag force.

We consider an electron fluid in two dimensions subject to a smooth random potential $V(\bm{r})$. In the collision-dominated regime, where the equilibration length due to electron-electron scattering is shorter than the correlation radius of the external potential, the electron system can be described by the hydrodynamic approach. The hydrodynamic equations consist of the continuity equation, the Navier-Stokes equation, and the entropy evolution equation:
\begin{eqnarray}
&&\hskip-.45cm
\partial_tn+\dv(n\bm{v})=0,\label{eq-continuity}\\
&&\hskip-.45cm
mn(\partial_t+\bm{v}\cdot\bm{\nabla})\bm{v}=\nonumber\\
&&-\bm{\nabla}P+n(\bm{F}-\bm{\nabla}U)-\frac{e}{c}[\bm{j}\times\bm{H}]+\bm{\nabla}\cdot\hat{\sigma},\label{eq-euler}\\
&&\hskip-.45cm
nT(\partial_t+\bm{v}\cdot\bm{\nabla})s=\dv (\kappa \boldsymbol{\nabla} T)-\varkappa \delta T .\label{eq-entropy}
\end{eqnarray}
Here $\bm{v}$ is the hydrodynamic velocity,  $m$ is the band mass of the electrons,  $n$ is the electron density, and $\bm{j}=n\bm{v}$ is the particle current density~\footnote{We assume parabolic band dispersion.}. The hydrodynamic approach assumes local thermal equilibrium so that the  pressure $P$ and entropy per particle $s$ may be expressed via the equation of state in terms of the temperature $T$ and density $n$.  The viscous stress tensor $\hat{\sigma}$ in Eq.~\eqref{eq-euler} has the standard form
\begin{equation}\label{eq-sigma}
\sigma_{ik}=\eta(\partial_kv_i+\partial_iv_k)
+(\zeta-\eta)\delta_{ik}\dv\bm{v},
\end{equation}
where $\eta$ and $\zeta$ are respectively the shear and bulk  viscosity coefficients. The derivatives $\partial_i$ and components of velocity field $v_i$ are taken with respect to the Cartesian coordinates in the plane. A uniform external magnetic field $\bm{H}$ produces a Lorentz force in Eq.~\eqref{eq-euler}, while $\bm{F}=e \bm{E}$ is the force exerted on an electron by a uniform external electric field $\bm{E}$ driving the flow. The potential $U=V+W$ in Eq.~\eqref{eq-euler} consists of the external potential $V$ and the self-consistent Coulomb potential $W$ caused by the density modulation of the electron liquid. The latter describes the long range part of the internal stresses arising in the liquid due to Coulomb interactions. The left hand side of the entropy evolution equation (\ref{eq-entropy}) describes convective transfer of heat. The first term in the right hand side of Eq.~\eqref{eq-entropy}   describes the heat flux relative to the electron liquid ($\kappa$ is the thermal conductivity of the liquid). The second term describes transfer of heat from the electron liquid to the substrate, with $\varkappa$ being the appropriate kinetic coefficient.

At zero current the particle density,  entropy per particle, temperature and pressure in the liquid are given by their equilibrium values, denoted by $n$, $s$, $T$, and $P$ respectively. In the presence of the current $\bm{j}$ these quantities acquire corrections; $\delta n$,  $\delta s$,  $\delta T$  and  $\delta P$.  The density of a uniform external force that must be exerted on the fluid in order to maintain a steady current can be found by averaging Eq.~\eqref{eq-euler} over space. At small current density it is linear in $\bm{j}$. Using the fact that internal stresses produce no net force on the liquid, to linear order in $\bm{j}$ we obtain from Eq.~\eqref{eq-euler}  the following expression for the external force density,
\begin{equation}\label{eq-F}
\bm{F}=\frac{1}{\langle n\rangle}\left\langle\delta n\bm{\nabla}V+\frac{e}{c}[\bm{j}\times\bm{H}]\right\rangle.
\end{equation}
Here $\langle\ldots\rangle$ denotes spatial averaging. The resistivity tensor $\hat\rho$ is related to $\bm{F}$ by $ e^2 \hat \rho \,  \langle \bm{j} \rangle =\bm{F}$. The second term in the right hand side of Eq.~\eqref{eq-F} describes the Hall component of the external force  and corresponds to the classical Hall resistivity $\rho_\perp=H/\langle n \rangle ec$. The determination of the longitudinal resistivity $\rho_\parallel$ amounts to evaluation of the first term.
Below we evaluate $\langle\delta n \bm{\nabla}V\rangle$ to the lowest nonvanishing (second) order in perturbation in the external  potential $V$.  Within this accuracy it is sufficient to determine the nonequilibrium correction to the density  $\delta n$  to linear order in $V$.

To this end, we  assume $\delta n/n\ll1$ and simplify Eqs.~\eqref{eq-continuity}--\eqref{eq-entropy} by retaining only terms linear in $V$ and $\bm{j}$. The continuity equation \eqref{eq-continuity} yields $\dv\bm{v}=-(\bm{j}\cdot\bm{\nabla}n)/n^2$.
By taking the divergence and curl of the linearized Eq.~\eqref{eq-euler}  we obtain
\begin{eqnarray}
\nabla^2\left[\delta P+ n \delta W+\frac{(\eta+\zeta)}{n^2} \,\bm{j}\cdot\bm{\nabla} n\right]=-\frac{en}{c}\bm{H}\cdot\curl\bm{v},\\
\eta\nabla^2\curl\bm{v}=\frac{e}{c}\bm{H}\,\left( \bm{j}\cdot\bm{\nabla}\ln n\right).
\end{eqnarray}
Excluding the curl of the velocity field we find
\begin{equation}\label{eq-deltaP}
\delta P + n \delta W=-\left[\frac{\eta+\zeta}{n}+\frac{n}{l^4_H\eta}\Delta^{-2}\right](\bm{j}\cdot\bm{\nabla}\ln n),
\end{equation}
where $\Delta$ denotes the Laplacian and we introduced the magnetic length $l_H=\sqrt{c/eH}$. The long range Coulomb forces described by $\delta W$ are determined by the induced density $\delta n$ via the Poisson equation (in the Fourier representation $\delta W_{\bm{q}}=2\pi e^2 \delta n_{\bm{q}}/q$). The nonequilibrium pressure, $\delta P$, depends not only on $\delta n$, but also on the temperature variation, $\delta T$; $\delta P = \left( \partial P /\partial n\right)_T \delta n + \left( \partial P /\partial T\right)_n \delta T $, where $\delta T$ may be determined by linearizing Eq.~\eqref{eq-entropy},
\begin{equation}\label{eq-deltaT}
(-\kappa\nabla^{2}+\varkappa)\delta T= - T\left( \frac{\partial s}{\partial n}\right)_T (\bm{j}\cdot\bm{\nabla}n).
\end{equation}
Here we used the relation  $\boldsymbol{\nabla}s =\left(  \frac{\partial s}{\partial n} \right)_T \boldsymbol{\nabla} n$.

Combining Eqs.~(\ref{eq-deltaP}), (\ref{eq-deltaT})  we obtain the nonequilibrium density in the Fourier representation,
\begin{equation}
\label{eq:delta_n_q}
\delta n_{\bm{q}}=  \frac{- i q\nu n_{\bm{q}}   \left( \bm{j} \cdot \bm{q}\right)}{q + q_{TF}}
  \left[ \frac{ T n^2 \left(  \frac{\partial s}{\partial n} \right)_T^2 }{\varkappa + \kappa q^2} + \frac{\eta + \zeta}{n^2} + \frac{1}{l_H^4 q^4 \eta}   \right].
\end{equation}
Here we introduced the thermodynamic density of states,  $\nu = \left( \frac{\partial n}{\partial \mu} \right)_T = n \left( \frac{\partial n }{\partial P}\right)_T$ and the inverse Thomas-Fermi  screening length in two dimensions, $q_{TF}= 2\pi e^2 \nu$.

Substituting Eq.~(\ref{eq:delta_n_q}) into Eq.~(\ref{eq-F}), and using the relation $n_{\bm{q}}=-q \, \nu V_{\bm{q}}/(q+q_{TF})$ from the linear screening theory, we obtain the following expression for the longitudinal resistivity
\begin{equation}
 \label{eq:rho_parallel}
 \rho_\parallel=\int \frac{d^2 q}{(2\pi)^2} \frac{ q^2 | n_{\bm{q}}|^2}{2 e^2} \left[    \frac{T \left(  \frac{\partial s}{\partial n} \right)_T^2 }{\varkappa + \kappa q^2}+ \frac{\eta + \zeta}{n^4}   + \frac{1}{n^2 (q l_H)^4 \eta} \right] .
\end{equation}
In the absence of heat transfer to the substrate, $\varkappa =0$, and at zero magnetic field this expression coincides with the result, Eq.~(6), of Ref.~\onlinecite{Andreev}. The magnetoresistance is described by the third term in the right hand side of Eq.~(\ref{eq:rho_parallel}). In contrast to the zero field resistance, which depends on both shear and bulk viscosities and the thermal conductivity of the liquid, the magnetoresistance depends only on the shear viscosity.  This enables extraction of viscosity of the  electron liquid  from magnetotransport measurements and in particular test applicabilty of the Fermi liquid theory in the regime of strong interaction when $r_s\gg1$. An alternative recently proposed approach involves Corbino disk device, which allows a determination of the viscosity of a quantum electron liquid from the \emph{dc} potential difference that arises between the inner and the outer edge of the disk in response to an oscillating magnetic flux~\cite{Polini-Corbino}.

Note that the dominant contribution to MR in Eq.~\eqref{eq:rho_parallel} arises from the long-range modulations of the equilibrium density. As a result, in the hydrodynamic regime this contribution exceeds the one arising from the dependence of the kinetic coefficients of the liquid on the magnetic field.
For example, in the Fermi-liquid regime
the hydrodynamic magnetoresistance in Eq.~\eqref{eq:rho_parallel} may be estimated as  $\rho_H \sim \frac{1}{e^2}  \frac{(\delta n)^2}{n^2} \frac{\xi^2}{R_c^2 k_F l}$, where $R_c \sim k_F l_H^2$ is the cyclotron radius, and  $\xi$ is the correlation length of density modulations. On the other hand, the  correction to, say the shear viscosity, due to the presence of a magnetic field is of the order $\delta \eta \sim - \eta l^2/R_c^2$, where $l $ is the mean free path due to electron-electron collisions.  The contribution to MR due to the magnetic field dependence of the viscosity is \textit{negative}~\cite{Alekseev} and may be estimated as $\delta \rho \sim \frac{1}{e^2} \frac{\delta n^2}{n^4 }  \frac{ k_F^3 l^3}{\xi^2 R_c^2}$. We thus see that in the hydrodynamic regime, $l\ll \xi$,  this correction is smaller than the hydrodynamic magnetoresistance in Eq.~\eqref{eq:rho_parallel};  $\frac{\delta \rho}{\rho_H} \sim \frac{l^4}{\xi^4}\ll1$.

Our derivation of Eq.~\eqref{eq:rho_parallel} elucidates the physical origin of the resistive force $\bm{F}$ in Eq.~\eqref{eq-F}. In the presence of the current the external potential $V$ moves relative to the liquid and produces linear in $V$ modulations in the pressure, temperature and viscous stresses that propagate the liquid.   The coupling of the  resulting modulation of the fluid density to the external potential produces a drag force \eqref{eq-F} in second order in $V$. This invokes an analogy with the problem of drag resistivity in the interactively coupled bilayers~\cite{Drag-Review}.  In the latter, thermal fluctuations of density in the passive layer induce modulations of the electron density in the active layer. Those, in turn, couple to the passive layer density fluctuations producing a drag force.  In the hydrodynamic regime both the intralayer and the drag resistivities are given by the sum of thermal and viscous contributions. As a result they have similar dependence on the viscosity and thermal conductivity of the fluid but differ in their respective temperature dependence~\cite{Hydro-drag,Drag-Crossover}.

For applications to particular systems it may be convenient to express the resistivity in Eq.~\eqref{eq:rho_parallel}  in terms of the doping potential.  To be specific let us consider an experimentally relevant setup in which  a doping layer is separated from the  two-dimensional electron system by a distance $d$. The average density of dopants is equal to the average density of electrons, $n$. For the spatially uncorrelated dopants the spectral power of the external random potential induced in the plane of the electron system is
\begin{equation} \label{eq:V_corr}
\langle|V_{\bm{q}}|^2\rangle=n\left(\frac{2\pi e^2}{q}\right)^2\exp(-2qd).
\end{equation}
For $d  \gg n^{-1/2}$ the screening is linear, and the equilibrium density modulation is related to the external potential as $n_{\bm{q}} =-\nu q V_{\bm{q}}/(q + q_{TF})$. Substituting  this relation and Eq.~\eqref{eq:V_corr} into Eq.~\eqref{eq:rho_parallel}, and  performing now momentum integrals one finds
\begin{equation}\label{eq:rho_Poisson}
\rho_\parallel=\frac{\rho_Q}{16\pi}\left[\frac{T}{nd^2\kappa}\left(\frac{\partial s}{\partial\ln n}\right)^2_T+\frac{3(\eta+\zeta)}{2n^3d^4} +\frac{4 \ln\left(\frac{L}{d}\right) }{n\eta l^4_H} \right],
\end{equation}
where $\rho_Q$ is quantum of resistance. When deriving the above result we assumed $q_{TF}d\gg1$, neglected external thermal losses $\varkappa\to0$, and used the thermodynamic identities $(\partial n/\partial T)_P=n^2(\partial s/\partial P)_T=n(\partial s/\partial\mu)_T$ leading to $\nu^{-2}(\partial n/\partial T)^2=(\partial s/\partial\ln n)^2_T$.  Note that for the disorder potential in Eq.~\eqref{eq:V_corr}
the momentum integral in the last term of Eq.~\eqref{eq:rho_parallel}
diverges logarithmically at small $q$. This divergence was cut off at a large spatial scale $L$ yielding the last term in Eq. \eqref{eq:rho_Poisson} that defines the margnetoresistance $\rho_H$.

The above results  depend on the microscopic properties of the electron fluid only via the magnitude and temperature dependence of kinetic coefficients and thermodynamic properties. A detailed microscopic theory for the temperature dependence of $\kappa$, $\eta$, $\zeta$ and $s$ of strongly correlated liquids in the quantum-nondegenerate regime has not been developed. In contrast, the Fermi-liquid regime has been studied extensively \cite{Abrikosov}. At temperatures below the Fermi energy, $T<E_F$, and assuming $r_s\sim1$, one readily finds: $\kappa\sim c_vnlv_F\sim E^2_F/T$, $\eta\sim mv_Fnl\sim nE^2_F/T^2$ and $c_v\sim T/E_F$, where $l=v_F\tau_{ee}$ is the electron-electron mean free path with $\tau^{-1}_{ee}\sim T^2/E_F$. The hydrodynamic description assumes a short mean free path, $l<d$, which restricts its range of applicability to  temperatures $T>T_1\sim E_F/\sqrt{k_Fd}$ with $k_Fd\gg1$. In this regime Eq.~\eqref{eq:rho_Poisson} yields the following estimate for the longitudinal resistivity
\begin{equation}
\frac{\rho_\parallel}{\rho_Q} \sim  \frac{1}{(k_Fd)^2}\left(\frac{T}{E_F}\right)^4+\frac{1}{(k_Fd)^4}\left(\frac{E_F}{T}\right)^2
+ \frac{\ln \left(\frac{L}{d} \right)}{(nl^2_H)^2}\left(\frac{T}{E_F}\right)^2  ,\label{eq:rho-FL}
\end{equation}
which is valid in a parametrically wide range of temperatures $T_1<T<E_F$. At the onset of the hydrodynamic regime, $T\sim T_1$, the viscous (second) term in Eq.~\eqref{eq:rho-FL} exceeds the thermal conductivity contribution (first term) by a parametrically large factor $k_Fd\gg1$ and resistivity is estimated to be of the order $\rho_\parallel/\rho_Q\sim1/(k_Fd)^3$.  In fact, viscosity governs resistivity in a window of temperatures $T_1<T<T_2$ and both contributions becomes of the same order in magnitude at $T_2\sim E_F/\sqrt[3]{k_Fd}$. At higher temperatures, $T_2<T<E_F$,  the thermal conductivity contribution becomes dominant.

\begin{figure}
\includegraphics[width=8cm]{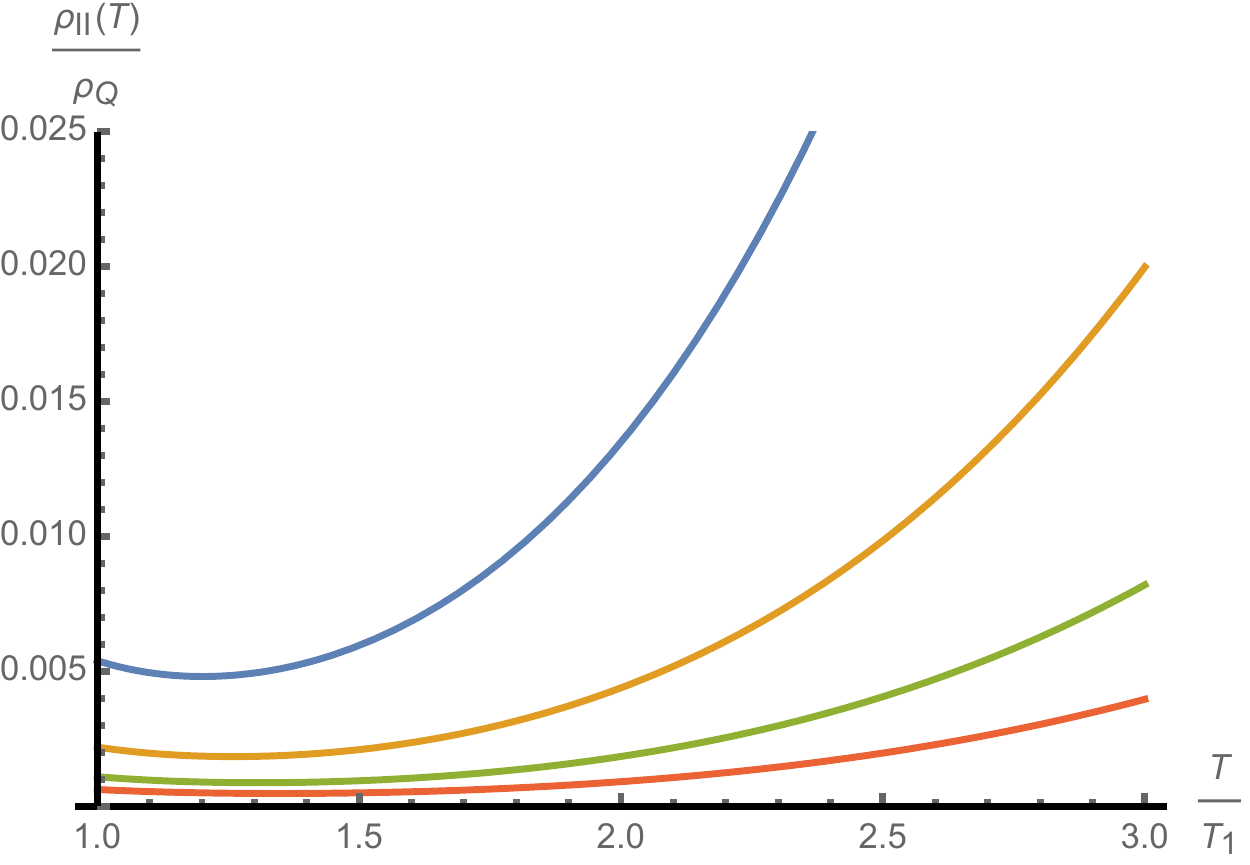}
\caption{Temperature dependence of the zero-field resistivity in Eq. \eqref{eq:rho-FL}  plotted for different values of the control parameter $k_Fd=6,8,10,12$, representing the curves from top to bottom respectively.}
\label{fig}
\end{figure}

In the context of  resistivity measurements carried out by different groups in various two-dimensional electron systems with moderate-to-strong interactions and deep in the metallic regime $\rho\ll\rho_Q$~\cite{Kravchenko,Spivak}, perhaps the most significant feature of Eq. \eqref{eq:rho-FL} is that it predicts a \textit{minimum} in the temperature dependence of the resistivity. At moderately high temperatures above $T_1$ the zero-field resistivity decreases first to a shallow minimum at around $T\sim T_2$ and then increases with $T$, as illustrated in Fig. \ref{fig}. We can further estimate the magnitude of the resistivity drop $\rho_\parallel(T_2)/\rho_\parallel(T_1)\sim 1/\sqrt[3]{k_Fd}$. We note that a minimum and a subsequent high-temperature rise of the resistivity is observed in many semiconductor devices with strongly correlated carriers, and usually attributed to the electron-phonon scattering, see for example Fig. 3(b) of Ref.~\onlinecite{Gao-PRL2005}. 

In summary, we have developed a hydrodynamic theory of  magnetoresistance of correlated electron systems subjected to long range disorder potential. Our theory applies in the collision-dominated regime, where the mean-free path due to electron-electron collisions is shorter than the correlation radius of the disorder potential and is not limited to the Fermi-liquid regime. The magnetoresistance is positive and depends only on the shear viscosity of the electron liquid. Assuming the Fermi liquid behavior we find that the zero-field resistivity, which is governed by the interplay of viscosity and thermal conductivity contributions displays a nonmonotonic temperature dependence. The description of the resistivity crossover to a low temperature collisionless regime of transport requires a separate investigation.

\textit{Acknowledgments}. We thank X. Gao and B. Shklovskii for useful discussions.
The work of  A. L. and H. X. was financially supported by the NSF Grants No. DMR-1606517 and ECCS-1560732. Support for this research at the University of Wisconsin-Madison was provided by the Office of the Vice Chancellor for Research and Graduate Education with funding from the Wisconsin Alumni Research Foundation. The work of A. A. was supported by the U.S. Department of Energy Office of Science, Basic Energy Sciences under Award No. DE-FG02-07ER46452.

\end{document}